\documentclass[twocolumn,aps,prl]{revtex4}
\usepackage{bm}
\usepackage{graphicx}
\newcommand{\calS}{{\mathcal S}}

\newcommand{\calZ}{{\mathcal Z}}
\newcommand{\calD}{{\mathcal D}}
\newcommand{\arcsinh}{{\rm arcsinh}}
\newcommand{\Det}{{\rm Det}}

\newcommand{\nbraket}[2]{\langle\bm{n}_#1|\bm{n}_#2\rangle}
\newcommand{\nrbraket}[2]{\left(\bm{n}_#1|\bm{n}_#2\right)}
\newcommand{\berry}[2]{\langle\bm{n}_{#1}|\partial_\tau\bm{n}_{#2}\rangle}
\begin{document}
\draft                                        
\preprint{}                                   
\title{ Spin wave spectrum of a
disordered double exchange model}
\author{ Takahiro Fukui}
 \affiliation{Department of Mathematical Sciences, 
Ibaraki University, Mito 310-8512, Japan}
\date{\today}
%---------------------------------------------------------------------
%                              Abstract
%---------------------------------------------------------------------
\begin{abstract}
A double exchange model with quenched disorder for conduction electrons
is studied by field theoretical methods. By using a path integral formalism
and replica techniques based on it, an ensemble-averaged spin wave dispersion 
of the localized spins is derived. 
It is shown that the spectrum of the spin wave has gaps 
at the multiple of the Fermi wavenumber of the conduction electrons 
in the presence of disorder, and hence, 
quenched disorder for electrons adds a striking effect to the dynamics of 
the localized spins. 
In the strong disorder limit, the present results suggest spin-glass like 
behavior due to the frustration of the exchange coupling.  
\end{abstract}

%\pacs{75.30.-m, 75.47.Lx, 05.30.-d, 11.10.-z}

\maketitle

The double exchange (DE) model \cite{Zen,AndHas,Gen,KubOha}
has been attracting much renewed interest 
due to the discovery of colossal magnetoresistence of 
perovskite manganites \cite{HWH,COK,JTM}.
These materials show a variety of phases, in which 
spin, charge, orbital, and lattice degrees of freedom 
and their interplay provide rich and complex behavior.
Although a lot of experimental and theoretical studies have been clarifying 
some aspects of complex physics of manganites, 
they still remain fascinating and incompletely understood phenomena \cite{Rev}.

It is believed that the DE mechanism is essential to the metallic
ferromagnetism of manganites. 
However, several authors have pointed out the importance of disorder,
which is inevitably included in the materials \cite{Var,SXS,AllAla,BMM}.
Especially, recent experiments have
reported anomalous behavior of spin wave spectrum, which cannot 
be explained by a simple DE theory \cite{HDC,DLM,DHZ,BHM}.
Motome and Furukawa \cite{MotFur}
have numerically studied DE model with quenched disorder and
found that the model successfully
describe the anomalies of spin wave dispersion.
They have suggested that the origin of the anomalies is the Friedel
oscillation. 

Motivated by their work, we study, in this paper, the effects of 
quenched disorder on the spin wave by field theoretical method.
Using the replica approach for the ensemble average over disorder
and integrating out the electrons,
we derive an effective action for the localized spins. 
The equation of motion derived from this action leads us to
the usual one for the Heisenberg ferromagnets but with 
$2k_F$ oscillating terms.   
To clarify our ideas in this paper, 
we restrict our discussions to one spatial dimension,
since disorder effects are more relevant and add a 
more striking effect in lower dimensions, 
as claimed by Motome and Furukawa.

%---------------------------------------------------------------------
%                       MODEL
%---------------------------------------------------------------------
The Hamiltonian we study is given by \cite{MotFur}
%---------------------------------------------------------------------
%   Hamiltonian
%---------------------------------------------------------------------
\begin{eqnarray}
H&=&-t\sum_{j}\sum_\sigma 
\left(
c_{j\sigma}^\dagger c_{j+1\sigma}+\mbox{h.c.}
\right)
-J_{H}\sum_j \bm{s}_j\cdot\bm{S}_j 
\nonumber\\&& %---------------------------------------------------!!!
+\sum_j\sum_\sigma\epsilon_jc_{j\sigma}^\dagger c_{j\sigma},
\label{Ham}%----------------------------------------------------------
\end{eqnarray}
where 
$\bm{s}_i
=\frac{1}{2}c_{i\sigma}^\dagger\bm{\sigma}_{\sigma\sigma'}c_{i\sigma'}$
denotes a spin operator of the conduction electron, 
$\bm{S}_j$ a localized spin which we treat as a classical object
$\bm{S}_j=S\bm{n}_j$ with a unit vector $\bm{n}_j$ defined by
$\bm{n}_j^{\rm t}=
(\sin\theta_j\cos\phi_j, \sin\theta_j\sin\phi_j, \cos\theta_j)$,
and 
$\epsilon_j$ 
is on-site disorder potential with the gaussian distribution
$P[\epsilon]=\exp\left[-\sum_j\epsilon_j^2/2g\right]$.

We start with the canonical DE model without disorder
to have a path integral formalism convenient for taking disorder into account.
The Hund's-rule coupling term can be written as
%---------------------------------------------------------------------
%   Hund-rule-coupling
%---------------------------------------------------------------------
$J_{H}\sum_j \bm{s}_j\cdot\bm{S}_j =
(J_{H}S/2)\bm{c}_j^\dagger\calS_j\bm{c}_j$,
where
$\bm{c}_j^\dagger=(c_{j\uparrow}^\dagger,c_{j\downarrow}^\dagger)$ and
%---------------------------------------------------------------------
%   \cal S_j
%---------------------------------------------------------------------
\begin{eqnarray}
\calS_j(\theta_j,\phi_j)\equiv
\left(
\begin{array}{ll}
\cos\theta_j&\sin\theta_j e^{-i\phi_j}\\
\sin\theta_j e^{i\phi_j}&-\cos\theta_j
\end{array}
\right) .
\label{HunRulMat}%----------------------------------------------------
\end{eqnarray}
The partition function is then represented by a path integral form
$\calZ=\int\calD\bm{c}\calD\bm{c}^\dagger e^{-S}$, where the action
is given by
%---------------------------------------------------------------------
%   partition function for the whole system
%---------------------------------------------------------------------
$S=\int_0^\beta d\tau
\left(
\sum_j \bm{c}_j^\dagger\partial_\tau\bm{c}_j
-H-\mu\sum_j\bm{c}_j^\dagger\bm{c}_j\right)$.
The matrix ${\cal S}_j$ in Eq. (\ref{HunRulMat}) is diagonalized 
by a {\it local} unitary matrix
%---------------------------------------------------------------------
%   unitary matrix
%---------------------------------------------------------------------
$
U_j(\tau)\equiv\calS_j(\theta_j/2,\phi_j)
$ as $U^\dagger_j(\tau)\calS_jU_j(\tau)=\sigma_3$.
Let us define a fermion field $\bm{\widetilde{c}}_i=U_i\bm{c}_i$
in the locally rotated frame. 
For a large $J_{H}$, only up-spin fermions survive, and  
we have an effective action described solely by 
fermions ${\widetilde c}_{j\uparrow}\equiv c_j$
in this limit,  
%---------------------------------------------------------------------
%   projected action
%---------------------------------------------------------------------
\begin{eqnarray}
S&=&\int_0^\beta d\tau \sum_j 
\Bigg[
c_j^\dagger\left(\partial_\tau+\berry{j}{j}\right) c_j
\nonumber\\&& %---------------------------------------------------!!!
-t\left(c_j^\dagger\nbraket{j}{{j+1}}c_{j+1}+\mbox{h.c.}\right)
-\mu c_j^\dagger c_j
\Bigg] .
\label{ProAct}%------------------------------------------------------
\end{eqnarray}
It should be noted that in this projection
the nonlocal transformation yields nontrivial
additional terms such as
$(U_j\partial_\tau U_j^\dagger)_{\uparrow\uparrow}$ 
and 
$(U_jU_{j+1}^\dagger)_{\uparrow\uparrow}$
in the canonical and hopping terms, respectively.
To describe these, we have introduced a bracket notation
%---------------------------------------------------------------------
%   coherent state
%---------------------------------------------------------------------
\begin{eqnarray}
\left(U_iU_j^\dagger\right)_{\uparrow\uparrow}
&=&
\cos\frac{\theta_i}{2}\cos\frac{\theta_j}{2}+
\sin\frac{\theta_i}{2}\sin\frac{\theta_j}{2}e^{-i(\phi_i-\phi_j)} 
\nonumber\\
&\equiv&
\nbraket{i}{j} ,
\end{eqnarray}
since this is actually equivalent to 
the inner product of the spin coherent state, 
%---------------------------------------------------------------------
%   Berry phase etc
%---------------------------------------------------------------------
$|\bm{n}_j\rangle=
\cos(\theta_j/2)|\uparrow\rangle+
\sin(\theta_j/2)e^{i\phi_j}|\downarrow\rangle$.
Therefore, 
$\berry{j}{j}\equiv A_j(\tau)$ 
can be interpreted as the Berry phase \cite{Ber},
playing an important role in the dynamics of the localized spins.

Assume that the localized spins are almost aligned to one direction,
$\nbraket{i}{j}\sim1$. 
In this case we introduce a small field variable
%---------------------------------------------------------------------
%   round braket
%--------------------------------------------------------------------- 
$\nrbraket{i}{j}=\nbraket{i}{j}-1$,
and then the action can be divided into two pieces 
$S=S_0+S_{\rm n}$, where
%---------------------------------------------------------------------
%   S_0 and S_n
%---------------------------------------------------------------------
\begin{eqnarray}
%\hspace*{-5mm}
S_0&=&\int_0^\beta \!\!\!d\tau\sum_j\left[
c_j^\dagger\partial_\tau c_j
-t\left(c_j^\dagger c_{j+1}+\mbox{h.c.}\right)-\mu c_j^\dagger c_j
\right],
\nonumber\\
%\hspace*{-5mm}
S_{\rm n}&=&\int_0^\beta \!\!\!d\tau\sum_j
\left[
c_j^\dagger A_jc_j
-t\left\{c_j^\dagger\nrbraket{j}{{j+1}} c_{j+1}+\mbox{h.c.}\right\}
\right] .
\nonumber\\ %-----------------------------------------------------!!!
\label{Stwo}%----------------------------------------------------------
\end{eqnarray}
These equations are basis of our perturbative calculations.
At the leading order, 
an effective action $W[\bm{n}]$ for the localized spins, 
defined by  
$e^{-W[\bm{n}]}=\langle e^{-S_{\rm n}}\rangle$,
is given by
%------------------------------------------------------------------
%   Effective action
%------------------------------------------------------------------
\begin{eqnarray}
W[\bm{n}]
=\int_0^\beta d\tau\sum_j
\left[
g_jA_j
-\frac{1}{2}J_{j}
\bm{n}_j\cdot\bm{n}_{j+1}
\right]
\label{EffAct}%-------------------------------------------------
\end{eqnarray}
except for some irrelevant constant terms, where
%------------------------------------------------------------------
%   g and J
%------------------------------------------------------------------
\begin{eqnarray}
&&
g_j\equiv\langle c_j^\dagger(\tau) c_j(\tau)\rangle,
\quad
J_{j}\equiv
t\langle c_j^\dagger(\tau) c_{j+1}(\tau)\rangle,
\label{CouCon}%-------------------------------------------------
\end{eqnarray}
and we have used the fact that
$|\nbraket{i}{j}|^2=\frac{1}{2}(1+\bm{n}_i\cdot\bm{n}_j)$,
namely, 
$\nrbraket{i}{j}+{\rm c.c.}\sim\frac{1}{2}(\bm{n}_i\cdot\bm{n}_j-1)$.
In the pure model under consideration, 
$g_j=1-x \equiv g_0$ 
and 
$J_{j}=2t\sum_{k>0}f(\varepsilon_k)\cos k \equiv J_0$ 
are uniform constants, 
where $x$ is the hole concentration, 
$\varepsilon_{k}=-2t\cos k-\mu$
is the transfer energy of electrons, and
$f(\varepsilon)=1/(e^{\beta\varepsilon}+1)$
is the Fermi distribution function. 
It should be noted that in Eq.  (\ref{EffAct}),
$\bm{n}_j\cdot\bm{n}_{j+1}\sim1$ is implicitly assumed.

The classical equation of motion for the localized spins ${\bm n}_j$
can be derived from the effective action Eq. (\ref{EffAct}): 
The variation of the action $W[\bm{n}]$ with respect to $\bm{n}_j$,
$\delta W/\delta \bm{n}_j(\tau)=0$,
leads us to 
%------------------------------------------------------------------
%   equation of motion
%------------------------------------------------------------------
\begin{eqnarray}
g_j\frac{d \bm{n}_j}{d t}
=
\bm{n}_j\times\left(J_{j-1}\bm{n}_{j-1}+J_{j}\bm{n}_{j+1}\right),
\label{EquMot}%-----------------------------------------------------
\end{eqnarray}
after the replacement $\tau\rightarrow -i t$.
In deriving this, the Berry phase is involved in the time derivative 
of the spin variable $\bm{n}_j$.
In the pure case, this equation is just the one for the Heisenberg
ferromagnets, giving dispersion relation
$\omega_k=(2J_0/g_0)(1-\cos k)$.

%-----------------------------------------------------------------------
%\section{Disorder effects}
%-----------------------------------------------------------------------

Now let us take account of disorder effects on the spin wave spectrum. 
Since the disorder potential in Eq. (\ref{Ham}) 
is invariant under the unitary transformation $U_j(\tau)$,
it gives rise to 
$H_{\rm d}=\sum_i\epsilon_ic_i^\dagger c_i$
in the projected action (\ref{ProAct}). 
We now introduce $m$ replicas to take quenched average over disorder.
Ensemble average  yields the following term
%------------------------------------------------------------------
%   emsenble-averaged S_d
%------------------------------------------------------------------
\begin{eqnarray}
S_{\rm d}
&=&
 -\frac{g}{2}\int_0^\beta \!\!\!d\tau\int_0^\beta \!\!\!d\tau'
\sum_i\sum_{a,b=1}^m
c_{ia}^\dagger(\tau) c_{ia}(\tau)c_{ib}^\dagger(\tau') c_{ib}(\tau')
\nonumber\\
&=&
\frac{g}{2}\sum_{a,b=1}^m
\sum_{\omega_n,\omega_n'}     
\sum_{k,k',q}                 
c_{k+q,a}^\dagger(i\omega_n)  c_{k,b}(i\omega_{n'})
\nonumber\\&&\times %-------------------------------------------------!!!
c_{k'-q,b}^\dagger(i\omega_{n'})c_{k',a}(i\omega_{n}) ,
\end{eqnarray}
where $a,b=1,2,\cdots,m$ denote replica species.
Although the fermions in the present theory are spinless (fully-polarized), 
this inter-replica coupling in $S_{\rm d}$ as well as the spin-fermion 
coupling in $S_{\rm n}$ are reminiscent of the theory of 
the spin density wave (SDW):
They are expected to play similar roles to the Coulomb interaction 
and phonons in SDW, respectively.
The wisdom obtained so far tells that
$2k_F$ oscillating mode of phonons makes a gap in the fermion sector,
where $k_F$ denotes the Fermi wavenumber.
On the analogy of this, we expect that the $2k_F$ mode of localized spins 
in $S_{\rm n}$ would be potentially involved with a gap formation 
of fermions, although we will treat 
$S_{\rm n}$ separately from $S_0+S_{\rm d}$, in what follows.
Based on these observations, 
let us first examine the mean field ground state of the fermions within
the action $S_0+S_{\rm d}$, and next compute the 
coupling constants of the localized spins, $g_j$ and $J_j$ 
in Eq. (\ref{CouCon}).
We have to mention here that $S_{\rm d}$ 
is nonlocal with respect to $\tau$, and therefore careful treatment
of the inter-replica interaction is needed.  Namely,
we divide the summation over the Matsubara frequencies into two parts:
\begin{widetext}%---------------------------------------------------!!!wide
%------------------------------------------------------------------
%   separate S_d into two parts
%------------------------------------------------------------------
\begin{eqnarray}
S_{\rm d}
&=&
\frac{g}{2}\sum_{a\ne b}
\sum_{\omega_{n}}  \sum_{\omega_{n_1},\omega_{n_2}} %%%\sum_{\omega_{n_2}}
\sum_{q,k,k'}         %%%\sum_{k'}\sum_q
c_{k+q,a}^\dagger(i\omega_{n_1}+i\omega_n)  c_{k,b}(i\omega_{n_1})
c_{k'-q,b}^\dagger(i\omega_{n_2}-i\omega_n) c_{k',a}(i\omega_{n_2})
\nonumber\\
&&+
\frac{g}{2}\sum_{a,b}
\sum_{\omega_{n}\ne0}\sum_{\omega_{n_1},\omega_{n_2}}
\sum_{q,k,k'}                %%%\sum_{k'}\sum_q
c_{k+q,a}^\dagger(i\omega_{n_1}+i\omega_n)c_{k,a}(i\omega_{n_1}) 
c_{k'-q,b}^\dagger(i\omega_{n_2}-i\omega_n) c_{k',b}(i\omega_{n_2}) .
\label{SdTwo}%-------------------------------------------------------
\end{eqnarray}
Here, $\omega_n=2\pi n/\beta$ is bosonic Matsubara frequency.
It should be noted that $a=b$ terms of 
the first line of the right hand side above
vanish identically due to the Fermi statistics. 
Introducing two kinds of auxiliary fields leads us to the action which is
bilinear in the Fermi fields,
%------------------------------------------------------------------
%   auxiliary fields
%------------------------------------------------------------------
\begin{eqnarray}
S_{\rm d}
&=&
\frac{1}{2g}\sum_{\omega_n}\sum_q\sum_{a\ne b}
Q_{q,ba}(i\omega_n)Q_{-q,ab}(-i\omega_n)+
\frac{1}{2g}\sum_{\omega_n\ne0}\sum_q
P_{q}(i\omega_n)P_{-q}(-i\omega_n)
\nonumber\\
&&+\frac{i}{2}\sum_{\omega_n}\sum_q
\sum_{\omega_{n_1}}\sum_k\sum_{a\ne b}
\Big[
Q_{q,ba}(i\omega_n)
c_{k+q,a}^\dagger(i\omega_{n_1}+i\omega_n)  c_{k,b}(i\omega_{n_1})
%\nonumber\\ && \hspace{35mm}%-----------------------------------!!!!!
+
Q_{-q,ab}(-i\omega_n)
c_{k-q,b}^\dagger(i\omega_{n_1}-i\omega_n) c_{k,a}(i\omega_{n_1}) 
\Big]
\nonumber\\
&&+\frac{i}{2}
\sum_{\omega_n\ne0}\sum_q
\sum_{\omega_{n_1}}\sum_k\sum_{a}
\Big[
P_{q}(i\omega_n)
c_{k+q,a}^\dagger(i\omega_{n_1}+i\omega_n)  c_{k,a}(i\omega_{n_1})
%\nonumber\\ && \hspace{35mm}%-------------------------------------!!!!!
+
P_{-q}(-i\omega_n)
c_{k-q,a}^\dagger(i\omega_{n_1}-i\omega_n) c_{k,a}(i\omega_{n_1}) 
\Big] .
%\nonumber\\ %-----------------------------------------------------!!!
\label{SdAux}%----------------------------------------------------
\end{eqnarray}
\end{widetext}%----------------------------------------------------!!!wide
%As a leading order approximation,
Let us now assume  a static ($i\omega_n=0$) 
saddle point solution with the following replica symmetric form
%------------------------------------------------------------------
%   SDW order
%------------------------------------------------------------------
\begin{eqnarray}
&&
Q_{q,ab}(i\omega_n)=2i\Delta\beta\delta_{n,0}\delta_{q,2k_F},
\label{OrdPar}%----------------------------------------------------
\end{eqnarray}
whereas $P_q(i\omega_n)=0$ since $i\omega_n=0$ is prohibited in the 
summation in Eq. (\ref{SdAux}).
This form of the mean field solution is similar to the one 
in the replica theory for spin glass models \cite{SheKir}. 
Although we have mentioned the similarity to the theory of SDW, 
the present replica coupling is repulsive.
Therefore, we cannot expect a stable mean field ground state
if the number of species $m$ is an integers $m\ge1$:
This is the reason we assume  an imaginary $Q$ in Eq. (\ref{OrdPar}).
Substituting Eq. (\ref{OrdPar}) into Eq. (\ref{SdAux}), we have
%------------------------------------------------------------------
%   S_d with SDW
%------------------------------------------------------------------
\begin{eqnarray}
S_{\rm d}=&&-
\sum_{a\ne b}\sum_{\omega_n}\sum_k
\Delta
\left[
c_{k+2k_F,a}^\dagger(i\omega_n)c_{k,b}(i\omega_n)+
\mbox{h.c}
\right]
\nonumber\\&& %-----------------------------------------------------!!!
-\frac{2m(m-1)}{g}\Delta^2.
\end{eqnarray}
Due to the assumption of imaginary $Q$, the $\Delta^2$ potential above 
has actually a wrong sign. 
Nevertheless, in the replica limit $m\rightarrow0$, we have 
a mean field solution, as we shall see momentarily.

Now let us determine the parameter $\Delta$ by minimizing the effective
action for $\Delta$.
To this end, we decompose the Fourier modes into several pieces,
%-----------------------------------------------------------------------
%   Fourier modes divided into 3 pieces
%-----------------------------------------------------------------------
\begin{eqnarray}\hspace{-3mm}
c_{j,a}(i\omega_n)
=\!\!\!\!\!\!\!
\sum_{|k|<k_F-\Lambda \atop |k|>k_F+\Lambda}e^{ikj}c_{k,a}(i\omega_n)+
\sum_{|k|<\Lambda}\bm{u}_{k,j}^\dagger\bm{c}_{k,a}(i\omega_n) ,
\end{eqnarray}
where 
$\bm{c}_{k,a}^\dagger\equiv
\left(c_{k+k_F,a}^\dagger,c_{k-k_F,a}^\dagger\right)$
and 
$\bm{u}_{k,j}^\dagger\equiv(e^{i(k+k_F)j},e^{i(k-k_F)j})$.
Then the fermion action becomes
\begin{widetext}%-----------------------------------------------------!!!wide
%-----------------------------------------------------------------------
%   S_0+S_d
%-----------------------------------------------------------------------
\begin{eqnarray}
S_0+S_{\rm d}
&=&
-\sum_{a,b}\sum_{\omega_n}
\Bigg[
\sum_{|k|<k_F-\Lambda\atop |k|>k_F+\Lambda}
c_{k,a}^\dagger(i\omega_n) 
\left(
i\omega_n-\varepsilon_{k}
\right)
c_{k,b}(i\omega_n)
%\nonumber\\ && %---------------------------------------------!!!!!
-\sum_{|k|<\Lambda}
\bm{c}_{k,a}^\dagger(i\omega_n)
\left[
D_k(i\omega_n)\delta_{ab}+M(1-\delta_{ab})
\right]
\bm{c}_{k,b}(i\omega_n)
\Bigg] ,
\nonumber\\%----------------------------------------------------------!!!
\end{eqnarray}
\end{widetext}%-------------------------------------------------------!!!wide
where
%-----------------------------------------------------------------------
%   definition of D and M
%-----------------------------------------------------------------------
\begin{eqnarray}
&&
D_k(i\omega_n)
\sim
\pmatrix{
i\omega_n-v_Fk & 0\cr
0& i\omega_n+v_Fk
},\quad
\nonumber\\&& %----------------------------------------------------------!!!
M=
\pmatrix{0&\Delta\cr \Delta&0} .
\end{eqnarray}
Here we have linearized the transfer energy of electrons as
$\varepsilon_{k\pm k_F}\sim \pm v_F k$ with 
$v_F\equiv 2t\sin k_F$. 
Note that the matrix $D_k\delta_{ab}+M(1-\delta_{ab})$ 
can be diagonalized in the replica space 
by a global unitary transformation,
giving $(m-1)$ eigenvalues $D_k-M$ and 1 eigenvalue $D_k-(1-m)M$.
Therefore, integrating out the Fermi fields yields 
$e^{-F(\Delta)}=\int\calD c\calD c^\dagger e^{-(S_0+S_{\rm d})}$,
where
%-----------------------------------------------------------------------
%   free energy
%-----------------------------------------------------------------------
\begin{eqnarray}
F(\Delta)&=&
-\frac{2m(m-1)\beta}{g}\Delta^2 
-(m-1) \ln \Det(D-M)
\nonumber\\&& %----------------------------------------------------!!!
-\ln \Det\left(D-(1-m)M\right) .
\end{eqnarray}
It is readily seen that $D_k(i\omega_n)-M$ has eigenvalues
$i\omega_n\pm E_k(\Delta)$ with 
$E_k(\Delta)=\sqrt{(v_Fk)^2+\Delta^2}$.
Then, the variation with respect to $\Delta$ gives
%-----------------------------------------------------------------------
%   variation of the free energy
%-----------------------------------------------------------------------
\begin{eqnarray}
&&
\frac{\partial F}{\partial \Delta}
=
m\Delta
\Bigg\{
\frac{4(1-m)}{g}
\nonumber\\&& %---------------------------------------------------!!!
-\frac{1}{v_F}
\ln\left[
\frac{\pi v_F\Lambda}{\Delta}
+\sqrt{1+\left(\frac{v_F\Lambda}{\Delta}\right)^2}\right]
+O(m)
\Bigg\},
\end{eqnarray}
where we have assumed that the temperature of the system 
is much lower than $\Delta$, i.e, $\beta\Delta\ll 1$.
This equation gives no solution in the normal case with $m\ge 1$,
since the inter-replica interaction is repulsive, as mentioned above.
However, it turns out that we obtain non trivial stable
solution in the replica limit $m\rightarrow0$, 
%-----------------------------------------------------------------------
%   stationary condition
%-----------------------------------------------------------------------
\begin{eqnarray}
\Delta=2v_F\Lambda
\frac{e^{\frac{4\pi v_F}{g}}}{e^{\frac{8\pi v_F}{g}}-1}
\sim
2v_F\Lambda e^{\frac{-4\pi v_F}{g}} .
\end{eqnarray}
This is due to the change of the sign of the $\Delta^2$ potential
at $m=1$.

Now we calculate the effective action Eq. (\ref{EffAct}) 
in the presence of disorder by using ensemble-averaged correlation 
functions in Eq. (\ref{CouCon}).
To this end, we need 
$\overline{\langle c_{i}^\dagger(\tau)c_{j}(\tau)\rangle}=
\lim_{m\rightarrow0}
\langle c_{i,1}^\dagger(\tau)c_{j,1}(\tau)\rangle$ in the replica approach. 
In order to calculate this in the replica-diagonal frame introduced 
previously, we have to note that the above correlation function 
is equivalent to 
$\lim_{m\rightarrow0}\frac{1}{m}\sum_{a}
\langle c_{i,a}^\dagger(\tau)c_{j,a}(\tau)\rangle$,
and therefore, invariant under the global transformation.
As we mentioned already, this transformation yields $m-1$ eigenvalues 
$D+M$ and 1 eigenvalue $D+(1-m)M$, the latter of which approaches the former
in the replica limit $m\rightarrow0$, and therefore,
this observation enables us to keep just one replica species, say,
$a=1$ with $D+M$ to calculate the above correlation function.
To simplify the notation, 
we will suppress the replica indices hereafter.

Based on 
%-----------------------------------------------------------------------
%   correlations in i,j-space
%-----------------------------------------------------------------------
\begin{eqnarray}
\langle
c_j(\tau)&&\hspace{-4mm}c_{j'}^\dagger(\tau)
\rangle
=
\sum_{\omega_n}
\Bigg[
\sum_{|k|<k_F-\Lambda\atop |k|>k_F+\Lambda}
e^{ik(j-j')}
\frac{-1}{i\omega_n-\varepsilon_k}
\nonumber\\&& %----------------------------------------------------!!!
+
\sum_{|k|<\Lambda}{\rm tr}
\bm{u}_{k,j'}\bm{u}_{k,j}^\dagger
\langle
\bm{c}_k(i\omega_n)\bm{c}_k^\dagger(i\omega_n)
\rangle
\Bigg] ,
\end{eqnarray}
and assume that $\Delta$ is much smaller than the temperature
under consideration, we end up with
%-----------------------------------------------------------------------
%   those for low-temperatures
%-----------------------------------------------------------------------
\begin{eqnarray}
g_j
&=&
g_0-g^{\rm osc}\cos(2k_Fj) ,
\nonumber\\
J_{j}
&=&
J_0-\delta J-J^{\rm osc}\cos[k_F(2j+1)] ,
\label{DisCouCon}%-----------------------------------------------------
\end{eqnarray}
where each term can be computed as
%-----------------------------------------------------------------------
%   oscilation-correction
%-----------------------------------------------------------------------
\begin{eqnarray}
g^{\rm osc}&=&2\sum_{0<k<\Lambda}\frac{\Delta}{E_k}
%\nonumber\\
=
\frac{\Delta}{\pi v_F}\arcsinh\left(\frac{\Lambda v_F}{\Delta}\right) ,
\nonumber\\
\delta J&=&2t\sin k_F\sum_{0<k<\Lambda}\sin k
\left(1-\frac{v_Fk}{E_k}\right) ,
\nonumber\\
J^{\rm osc}&=&2t\sum_{0<k<\Lambda}\cos k\frac{\Delta}{E_k} .
\label{ConCouCons}%-----------------------------------------------------
\end{eqnarray}
Thus we have derived a classical ferromagnetic spin-wave action 
(\ref{EffAct}) with Eqs. (\ref{DisCouCon}) and (\ref{ConCouCons}), 
which yields the equation of motion
(\ref{EquMot}). Eq. (\ref{DisCouCon}) tells us 
that {\it nonmagnetic impurities of conduction electrons} 
give rise to a $2k_F$ oscillating component to the 
{\it exchange coupling of the localized spins}.
This results in {\it the gap opening in the spin wave dispersion} 
$\omega_q$ at 
$q=k_F, 2k_F,\cdots\mbox{mod}(\pi)$, which are controlled by 
{\it the Fermi wavenumber $k_F$ of conduction electrons}.   
%-----------------------------------------------------------------------
%   Fig.1
%-----------------------------------------------------------------------
\begin{figure}
\includegraphics[width=7cm]{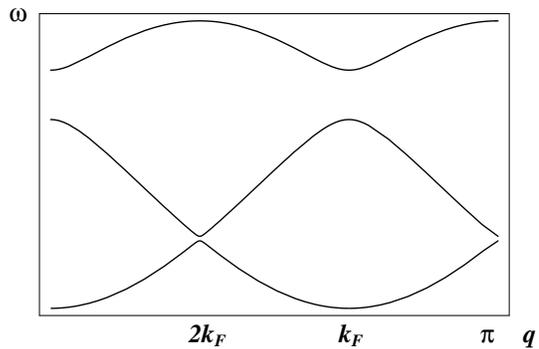}
\caption{Schematic illustration of the spin wave dispersion for 
the $x=1/3$ concentration case in the periodic zone scheme.
Gaps open at $q=k_F=2/3\pi$ and $2k_F=1/3\pi$ (mod $\pi$).}
\label{f:Dis}%----------------------------------------------------------
\end{figure}
We show in Fig. \ref{f:Dis} a schematic spin wave spectrum 
for a $x=1/3$ doping case.
As expected from the exchange coupling with
the $2k_F$ oscillation, there appears a gap at
$q=(2/3)\pi$. In addition, the dispersion relation has a small gap
at $q=2k_F=(1/3)\pi \mbox{~mod~} \pi$. 
In general, when the Fermi wavenumber is commensurate, $k_F=(m/n)\pi$,
there appears $n$ gaps in the spin wave dispersion.
It should be noted that 
$2k_F$ oscillation is included
not only in the exchange coupling but also in the Berry phase term,
both of which determine the size of gaps.
As far as we have studied the equation of motion numerically, 
$g_j$ tends to make the gap at $q=k_F$ larger and the other gaps
smaller. Details will be published elsewhere.

In a strong disorder limit, we expect relatively large $\Delta$
of order $\Lambda$.  
Although the present formalism is a weak disorder approach, 
it may be suggestive to consider the case
$\Delta\rightarrow \infty$. Setting $\Lambda\sim k_F$, we have
%-----------------------------------------------------------------------
%  large gap limit
%-----------------------------------------------------------------------
$g_j\sim 2(1-x)\sin^2(k_Fj)$ and 
$J_{j}\sim\frac{v_F}{\pi}\sin[k_F(j+1)]\sin(k_Fj)$.
These equations tell that the exchange coupling could be frustrated,
which may be responsible for the glass phase of the manganites
observed actually by experiments \cite{AUT}.
However, in this case, quantum treatments of the localized spins would
be required.

So far we have studied a spin wave dispersion in one dimension, 
which has gaps for {\it any disorder strength g}.
In higher dimensions, there exists a critical value $g_{\rm c}(>0)$,
below which the dispersion has no gaps. 
Details will be published elsewhere. 

%\begin{acknowledgments}

The author would like to thank Y. Motome and M. Yokoyama 
for valuable discussions.
This work was supported in part by Grant-in-Aid for scientific Research
from JSPS.
%\end{acknowledgments}

%--------------------------------------------------------------------
%   references
%--------------------------------------------------------------------

\end{document}